\documentclass{emulateapj}
\usepackage{amsbsy}

\newcommand{\vc}[1]{{\boldsymbol #1}}
\newcommand{\de}{\mathrm{d}}
\newcommand{\dpa}{\partial}

\newcommand{\nab}{\vc{\nabla}}
\newcommand{\ee}{\mathrm e}
\DeclareMathSymbol{\varOmega}{\mathord}{letters}{"0A}
\DeclareMathSymbol{\varSigma}{\mathord}{letters}{"06}
\DeclareMathSymbol{\varPsi}{\mathord}{letters}{"09}

\newcommand{\Eq}[1]{equation (\ref{#1})}
\newcommand{\Eqs}[2]{equations (\ref{#1}) and~(\ref{#2})}
\newcommand{\Eqss}[2]{equations (\ref{#1})--(\ref{#2})}

\newcommand{\Sec}[1]{Sect.~\ref{#1}}

\newcommand{\Fig}[1]{Fig.~\ref{#1}}
\newcommand{\Figs}[2]{Figs.~\ref{#1} and \ref{#2}}

\newcommand{\Tab}[1]{Table \ref{#1}}

\shorttitle{Dust Sedimentation and Kelvin-Helmholtz Turbulence}
\shortauthors{Johansen et al.}

\begin{document}


\title{Dust Sedimentation and Self-Sustained Kelvin-Helmholtz Turbulence in
Protoplanetary Disk Mid-Planes. I. Radially symmetric simulations.}


\author{Anders Johansen, Thomas Henning and Hubert Klahr}
\affil{Max-Planck-Institut f\"ur Astronomie, K\"onigstuhl 17, 69117 Heidelberg,
Germany}
\email{johansen@mpia.de}




\begin{abstract}

We perform numerical simulations of the Kelvin-Helmholtz instability in the
mid-plane of a protoplanetary disk. A two-dimensional corotating slice in the
azimuthal--vertical plane of the disk is considered where we include the
Coriolis force and the radial advection of the Keplerian rotation flow.  Dust
grains, treated as individual particles, move under the influence of friction
with the gas, while the gas is treated as a compressible fluid. The friction
force from the dust grains on the gas leads to a vertical shear in the gas
rotation velocity. As the particles settle around the mid-plane due to gravity,
the shear increases, and eventually the flow becomes unstable to the
Kelvin-Helmholtz instability. The Kelvin-Helmholtz turbulence saturates when
the vertical settling of the dust is balanced by the turbulent diffusion away
from the mid-plane. The azimuthally averaged state of the self-sustained
Kelvin-Helmholtz turbulence is found to have a constant Richardson number in
the region around the mid-plane where the dust-to-gas ratio is significant.
Nevertheless the dust density has a strong non-axisymmetric component.
We identify a powerful clumping mechanism, caused by the dependence of the
rotation velocity of the dust grains on the dust-to-gas ratio, as the source of
the non-axisymmetry. Our simulations confirm recent findings that the critical
Richardson number for Kelvin-Helmholtz instability is around unity or larger,
rather than the classical value of 1/4.

\end{abstract}





\section{INTRODUCTION}

One of the great unsolved problems of planet formation is how to form
planetesimals, the kilometer-sized precursors of real planets
\citep{Safronov1969}. At this size solid bodies in a protoplanetary disk can
attract each other through gravitational two-body encounters, whereas gravity
is insignificant between smaller bodies. Starting from micrometer-sized dust
grains, the initial growth is caused by the random Brownian motion of the
grains \citep[e.g.][see \cite{Henning+etal2006} for a
review]{BlumWurm2000,DullemondDominik2005}. The vertical component of the
gravity from the central object causes the gas in the disk to be stratified
with a higher pressure around the mid-plane. Even though the dust grains do not
feel this pressure gradient, the strong frictional coupling with the gas
prevents small grains from having any significant vertical motion relative to
the gas. However, once the grains have coagulated to form pebbles with sizes of
a few centimeters, the solids are no longer completely coupled to the gas
motion.  They are thus free to fall, or sediment, towards the mid-plane of the
disk. The increase in dust density opens a promising way of forming
planetesimals by increasing the local dust density around the mid-plane of the
disk to values high enough for gravitational fragmentation of the dust layer
\citep{Safronov1969,GoldreichWard1973}.

There are however two major unresolved problems with the gravitational
fragmentation scenario. Any global turbulence in the disk causes the dust
grains to diffuse away from the mid-plane, and thus the dust density is kept at
values that are too low for fragmentation. A turbulent $\alpha$-value of
$10^{-4}$ is generally enough to prevent efficient sedimentation towards the
mid-plane \citep{WeidenschillingCuzzi1993}, whereas the $\alpha$-value due to
magnetorotational turbulence
\citep{BalbusHawley1991,Brandenburg+etal1995,Hawley+etal1995,Armitage1998} is
from a few times $10^{-3}$ (found in local box simulations with no imposed
magnetic field) to 0.1 and higher
(in global disk simulations). The presence of a magnetically dead zone around
the disk mid-plane \citep{Gammie1996,Fromang+etal2002,Semenov+etal2004} may not
mean that there is no turbulence in the mid-plane, as other instabilities may
set in and produce significant turbulent motion \citep{Li+etal2001,
KlahrBodenheimer2003}. The magnetically active surface layers of the disk can
even induce enough turbulent motion in the mid-plane to possibly prevent
efficient sedimentation of dust \citep{FlemingStone2003}. The presence of a
dead zone may actually {\it in itself} be a source of turbulence. The sudden
fall of the accretion rate can lead to a pile up of mass in the dead zone,
possibly igniting the magnetorotational instability in bursts
\citep{Wuensch+etal2005} or a Rossby wave instability
\citep{VarniereTagger2005}.

The second major problem with the gravitational fragmentation scenario is that
even in the absence of global disk turbulence, the dust sedimentation may in
itself destabilize the disk. Protoplanetary disks have a radial pressure
gradient, because the temperature and the density fall with increasing radial
distance from the central object, so the gas rotates at a speed that is
slightly below the Keplerian value. The dust grains feel only the gravity and
want to rotate purely Keplerian. Close to the equatorial plane of the disk,
where the sedimentation of dust has increased the dust-to-gas ratio to unity or
higher, the gas is forced by the dust to orbit at a higher speed than far away
from the mid-plane where the rotation is still sub-Keplerian. Thus there is a
vertical dependence of the gas rotation velocity. Such shear flow can be
unstable to the Kelvin-Helmholtz instability (KHI), depending on the
stabilizing effect of vertical gravity and density stratification. A necessary
criterion for the KHI is that the energy required to lift a fluid
parcel of gas and dust vertically upwards by an infinitesimal distance is
available in the relative vertical motion between infinitesimally close
parcels \citep{Chandrasekhar1961}. The turbulent motions resulting from the KHI
are strong enough to puff up the dust layer and prevent the formation of an
infinitesimally thin dust sheet around the mid-plane of the disk
\citep{Weidenschilling1980,WeidenschillingCuzzi1993}.

Modifications to the gravitational fragmentation scenario have been suggested
to overcome the problem of Kelvin-Helmholtz turbulence. \citeauthor{Sekiya1998}
(1998, hereafter referred to as S98) found that if the mid-plane of the disk is
in a state of constant Richardson number, as expected for small grains whose
settling time is long compared to the growth rate of the KHI, then an increase
in the global dust-to-gas ratio can lead to the formation of a high density
dust cusp very close to the mid-plane of the disk, reaching potentially a
dust-to-gas ratio of 100 already at a global dust-to-gas ratio that is 10 times
the canonical interstellar value of $0.01$. The appearance of a superdense dust
cusp in the very mid-plane has been interpreted by \cite{YoudinShu2002} as an
inability of the gas (or of the KHI) to move more mass than its own away from
the mid-plane.  As a source of an increased value of the global dust-to-gas
ratio, \cite{YoudinShu2002} suggest that the dust grains falling radially
inwards through the disk pile up in the inner disk. A slowly growing radial
self-gravity mode in the dust density has also been suggested as the source of
an increased dust-to-gas ratio at certain radial locations
\citep{Youdin2005a,Youdin2005b}. Trapping dust boulders in a turbulent flow is
a mechanism for avoiding the problem of self-induced Kelvin-Helmholtz
turbulence altogether
\citep{BargeSommeria1995,KlahrHenning1997,HodgsonBrandenburg1998,Chavanis2000,Johansen+etal2004}.
If the dust can undergo a gravitational fragmentation locally, because the
boulders are trapped in features of the turbulent gas flow such as vortices or
high-pressure regions, then there is no need for an extremely dense dust layer
around the mid-plane.  \citet*{Johansen+etal2006} found that meter-sized dust
boulders are temporarily trapped in regions of slight gas overdensity in
magnetorotational turbulence, increasing the dust-to-gas ratio locally by up to
two orders of magnitude. They estimate that the dust in such regions should
have time to undergo gravitational fragmentation before the high-pressure
regions dissolve again.  \cite{FromangNelson2005}, on the other hand, find that
vortices can even form in magnetorotationally turbulent disks, keeping dust
boulders trapped for hundreds of disk rotation periods. The KHI cannot operate
inside a vortex because there is no radial pressure gradient, and thus no
vertical shear, in the center of the vortex \citep{KlahrBodenheimer2006}.

From a numerical side it has been shown many times that a pure shear flow,
i.e.\ one that is not explicitly supported by any forces, is unstable, both
with magnetic fields \citep{Keppens+etal1999, KeppensToth1999} and without
\citep{BalbusHawleyStone1996}. But the key point here is that the vertical
shear formed in a protoplanetary disk is due to the sedimentation of dust, and
that the shear is able to regenerate as the dust falls down again, thus keeping
the flow unstable to KHI. The description of the full non-linear outcome of
such a system requires numerical simulations that include dust that can move
relative to the gas.

Linear stability analysis of dust-induced shear flows in protoplanetary disks
have been performed for simplified physical conditions
\citep{SekiyaIshitsu2000}, but also with Coriolis forces and Keplerian shear
included \citep{IshitsuSekiya2002,IshitsuSekiya2003}. Recently
\citeauthor{GomezOstriker2005} (2005, hereafter referred to as GO05) took an
approach to include the dust into their numerical simulations of the
Kelvin-Helmholtz instability by having the dust grains so extremely
well-coupled to the gas that they always move with the instantaneous velocity
of the gas. This is indeed a valid description of the dynamics of tiny dust
grains. However, the strong coupling to the gas does not allow the dust grains
to fall back towards the mid-plane. Thus the saturated state of the
Kelvin-Helmholtz turbulence can not be reached this way.

In this paper we present computer simulations where we have let the dust
grains, represented by particles each with an individual velocity vector and
position, move relative to the gas. This allows us to obtain a state of
self-sustained Kelvin-Helmholtz turbulence from which we can measure quantities
such as the diffusion coefficient and the maximum dust density. A better
knowledge of these important characteristics of Kelvin-Helmholtz turbulence is
vital for our understanding of planet formation.

\section{DYNAMICAL EQUATIONS}\label{ch:dynamical_equations}

We start by introducing the dynamical equations that we are going to solve for
the gas and the particles.

We consider a protoplanetary disk as a plane rotating with the Keplerian
frequency $\varOmega_0$ at a
distance $r=r_0$ from a protostellar object. The plane is oriented so that only
the azimuthal and vertical directions (which we name $y$ and $z$, respectively)
of the disk are treated. The absence of the radial $x$-direction means that the
Keplerian shear is ignored. The onset of the KHI is very likely to be affected
by the presence of radial shear, since the unstable modes of the KHI are
non-axisymmetric and therefore will be sheared out, but we believe the problem
of Kelvin-Helmholtz turbulence in protoplanetary disks to be rich enough to
allow for such a simplification as a first approach. There is of course the
risk that the nature of the instability could change significantly with the
inclusion of Keplerian shear \citep{IshitsuSekiya2003}, but on the other hand,
the results that we present here regard mostly the dynamics of dust particles
in Kelvin-Helmholtz turbulence, and we expect the qualitative results to hold
even with the inclusion of the Keplerian shear.

As a dynamical solver we use the Pencil Code, a finite difference code that
uses sixth order centered derivatives in space and a third order Runge-Kutta
time integration scheme\footnote{The code, including modifications made for the
current work, is available at\\
\url{http://www.nordita.dk/software/pencil-code/}}. See \cite{Brandenburg2003}
for details on the numerical schemes and test runs.

\subsection{Gas Equations}\label{ch:gas_equations}

The three components of the equation of motion of the gas are
\begin{eqnarray}
  \frac{\dpa u_x}{\dpa t} + (\vc{u}\cdot\nab)u_x
      &=& 2 \varOmega_0 u_y \nonumber \\
      & & \quad - \frac{1}{\gamma} c_{\rm s} \varOmega_0 \beta
      - \frac{\epsilon}{\tau_{\rm f}}\left(u_x-w_x\right) \, ,
  \label{eq:eqmot1}\\
  \frac{\dpa u_y}{\dpa t} + (\vc{u}\cdot\nab)u_y
      &=& -\frac{1}{2} \varOmega_0 u_x \nonumber \\
      & & \quad - \frac{1}{\rho} \frac{\dpa P}{\dpa y}
      - \frac{\epsilon}{\tau_{\rm f}}\left(u_y-w_y\right) \, ,
  \label{eq:eqmot2}\\
  \frac{\dpa u_z}{\dpa t} + (\vc{u}\cdot\nab)u_z
      &=& -\varOmega_0^2 z - \frac{1}{\rho} \frac{\dpa P}{\dpa z}
      - \frac{\epsilon}{\tau_{\rm f}}\left(u_z-w_z\right) \, .
  \label{eq:eqmot3}
\end{eqnarray}
Here $\vc{u}=(u_x,u_y,u_z)$ denotes the velocity field of the gas measured
relative to the Keplerian velocity. We explain now in some detail the terms
that are present on the right-hand-side of \Eqss{eq:eqmot1}{eq:eqmot3}. The
$x$- and $y$-components of the equation of motion contain the Coriolis force
due to the rotating disk, to ensure that there is angular momentum
conservation. Since velocities are measured relative to the Keplerian shear
flow, the advection of the rotation flow by the radial velocity component has
been added to the azimuthal component of the Coriolis force, changing the
factor $-2$ to $-1/2$ in \Eq{eq:eqmot2}.  We consider local pressure gradient
forces only in the azimuthal and vertical directions, whereas there is a
constant global pressure gradient force in the radial direction.  The global
density is assumed to fall radially outwards as $\dpa\ln\rho/\dpa \ln
r=\alpha$, where $\alpha$ is a constant. Assuming for simplicity that the
density decreases isothermally, we can write the radial pressure gradient force
as
\begin{equation}
  -\frac{1}{\rho} \frac{\dpa P}{\dpa r} =
      -\frac{1}{\gamma} c_{\rm s}^2 \frac{\dpa \ln \rho}{\dpa r} \, .
\end{equation}
Here $\gamma=5/3$ is the ratio of specific heats and $c_{\rm s}$ is the constant
sound speed. Rewriting this expression and using the isothermal disk expression
$H=c_{\rm s}/\varOmega_0$, we arrive at the expression
\begin{equation}\label{eq:radpresg}
  -\frac{1}{\rho} \frac{\dpa P}{\dpa r} =
      -\frac{1}{\gamma} \frac{H}{r} \frac{\dpa \ln \rho}{\dpa \ln r} c_{\rm s}
      \varOmega_0 \, .
\end{equation}
We then proceed by defining the dimensionless disk parameter
$\beta\equiv\frac{H}{r} \frac{\dpa \ln \rho}{\dpa \ln r}$, where $H/r$ is the
scale-height to radius ratio of the disk. This parameter can be assumed to be a
constant for a protoplanetary disk. Using \Eq{eq:radpresg} and the definition
of $\beta$ leads to the global pressure gradient term in \Eq{eq:eqmot1}. We
use throughout this work a value of $\beta=-0.1$.

The ratio between the pressure gradient force $\Delta g$ and two times the
solar gravity,
\begin{equation}\label{eq:eta_def}
  \eta = \frac{\Delta g}{2 g} = \frac{-\frac{1}{\rho} \frac{\dpa P}{\dpa r}}{2
      \varOmega_0^2 r} \, ,
\end{equation}
is often used to parameterize sub-Keplerian disks \citep{Nakagawa+etal1986}.
Assuming again an isothermally falling density, \Eq{eq:eta_def} can be written
as
\begin{equation}
  \eta = -\frac{1}{2} \frac{1}{\gamma} \left( \frac{H}{r} \right)^2 \frac{\dpa
  \ln \rho}{\dpa \ln r} \, .
\end{equation}
The connection between our $\beta$ and the more widely used $\eta$ is then
\begin{equation}\label{eq:eta_beta}
  \eta = -\frac{1}{2} \frac{1}{\gamma} \frac{H}{r} \beta \, .
\end{equation}

The last term in \Eqss{eq:eqmot1}{eq:eqmot3} is the friction force that the
dust particles exert on the gas. We discuss this in further detail in
\Sec{ch:dust_equations} below.  Here the dust velocity field
$\vc{w}=(w_x,w_y,w_z)$ is a map of the particle velocities onto the grid. To
stabilize the finite difference numerical scheme of the Pencil Code, we use a
sixth-order momentum-conserving hyperviscosity
\citep[e.g.][]{BrandenburgSarson2002,HaugenBrandenburg2004,JohansenKlahr2005}.
Hyperviscosity has the advantage over regular second-order viscosity in that it
has a huge effect on unstable modes at the smallest scales of the simulation,
while leaving the largest scales virtually untouched.

\subsection{Mass and Energy Conservation}

The conservation of mass, given by the logarithmic density $\ln \rho$, and
entropy, $s$, is expressed in the continuity equation and the heat equation,
\begin{eqnarray}
  \frac{\dpa \ln \rho}{\dpa t} + (\vc{u}\cdot\nab) \ln \rho
      &=& - \nab \cdot \vc{u} \, ,
  \label{eq:conteq}\\
  \frac{\dpa s}{\dpa t} + (\vc{u}\cdot\nab) s &=& 0 \, .
  \label{eq:heateq}
\end{eqnarray}
The advection of the global density gradient due to any radial velocity has
been ignored, as well as viscous heating of the gas. We calculate the pressure
gradient force in \Eqss{eq:eqmot1}{eq:eqmot3} by rewriting the vector term as
\begin{equation}
  -\rho^{-1}\nab P = -c_{\rm s}^2(\nab s/c_{\rm p}+\nab\ln\rho)
\end{equation}
and using the ideal gas law expression
\begin{equation}
  c_{\rm s}^2 = \gamma \frac{P}{\rho}
      = c_{\rm s0}^2 \exp \left[ \gamma s/c_{\rm p}
      + (\gamma{-}1)\ln\frac{\rho}{\rho_0} \right] \, .
\end{equation}
The two constants $c_{\rm s0}$ and $\rho_0$ are integration constants from the
integration of the first law of thermodynamics. We have chosen the integration
constants such that $s=0$ when $c_{\rm s}=c_{\rm s0}$ and $\rho=\rho_0$.  To
allow for gravity waves, we must use the perfect gas law, rather than a simple
polytropic equation of state.  We stabilize the continuity equation and the
entropy equation by using a 5th order upwinding scheme
\citep[see][]{Dobler+etal2006} for the advection terms in
\Eqs{eq:conteq}{eq:heateq}.

\subsection{Dust Equations}\label{ch:dust_equations}

\begin{deluxetable*}{llrrrrrr}

  \tablewidth{0pt}
  \tablecaption{Run parameters}
  \tablehead{\colhead{Run} & \colhead{$\varOmega_0 \tau_{\rm f}$} &
      \colhead{$\epsilon_0$} & \colhead{$L_y\times L_z$} & \colhead{$N_y\times
      N_z$} & \colhead{$N$} & \colhead{$N_1$} & \colhead{$\Delta t/\varOmega_0^{-1}$}}

  \startdata

    A   & $0.02$ & $0.01$ &$0.40\times0.20$ & $256\times128$ &
        400,000  & $12.2$ & $200$ \\
    B   & $0.10$  & $0.01$ &$0.40\times0.20$ & $256\times128$ &
        400,000  & $12.2$ & $200$ \\
    C   & $1.00$  & $0.01$ &$0.10\times0.05$ & $256\times128$ &
        400,000  & $12.2$ &  $80$ \\
    Be2  & $0.10$  & $0.02$ &$0.40\times0.20$ & $256\times128$ &
        400,000  & $12.2$ & $100$ \\
    Be5  & $0.10$  & $0.05$ &$0.40\times0.20$ & $256\times128$ &
        400,000  & $12.2$ & $100$ \\
    Be10 & $0.10$  & $0.10$ &$0.40\times0.20$ & $256\times128$ &
        400,000  & $12.2$ & $100$ \\
    Br512 & $0.10$  & $0.01$ &$0.40\times0.20$ & $512\times256$ &
        1,600,000  & $12.2$ & $100$

  \enddata

  \tablecomments{Col. (1): Name of run. Col. (2): Stokes number.
  Col. (3): Global dust-to-gas ratio. Col. (4): Size of simulation box.
  Col. (5): Grid resolution. Col. (6): Number of particles.
  Col. (7): Number of particles per grid point.
  Col. (8): Total time of run in units of $\varOmega_0^{-1}$}
  \label{t:parameters}
\end{deluxetable*}
The dust grains are treated as individual particles moving on the top of the
grid. Therefore they have no advection term in their equation of motion, whose
components are
\begin{eqnarray}
  \frac{\de v_x^{(i)}}{\de t} &=& 2 \varOmega_0 v_y^{(i)}
      - \frac{1}{\tau_{\rm f}} \left(v_x^{(i)}-u_x\right) \, ,
  \label{eq:eqmotp1}\\
  \frac{\de v_y^{(i)}}{\de t} &=& -\frac{1}{2} \varOmega_0 v_x^{(i)}
      - \frac{1}{\tau_{\rm f}} \left(v_y^{(i)}-u_y\right) \, ,
  \label{eq:eqmotp2}\\
  \frac{\de v_z^{(i)}}{\de t} &=& -\varOmega_0^2 z^{(i)}
      - \frac{1}{\tau_{\rm f}} \left(v_z^{(i)}-u_z\right) \, .
  \label{eq:eqmotp3}
\end{eqnarray}
The index i runs in the interval $i=1\ldots N$, where $N$ is the number of
particles that are considered. The last terms in \Eqss{eq:eqmotp1}{eq:eqmotp3}
is the friction force. The friction force is assumed to be proportional to the
velocity difference between dust and gas with a characteristic braking-down
time-scale of $\tau_{\rm f}$, called the friction time. To conserve the total
momentum, the dust must affect the gas by an oppositely directed friction force
with friction time $\tau_{\rm f}/\epsilon$, as included in the last terms of
\Eqss{eq:eqmot1}{eq:eqmot3}. Here $\epsilon$ is the local dust-to-gas mass
ratio $\rho_{\rm d}/\rho$. The dust density $\rho_{\rm d}$ at a grid point is
calculated by counting the number of particles within a grid cell volume around
the point and multiplying by the mass density $\tilde{\rho}_{\rm d}$ that each
particle represents. The mass density per particle depends on the number of
particles and on the assumed average dust-to-gas ratio $\epsilon_0$ as
$\tilde{\rho}_{\rm d}=\epsilon_0 \rho/N_1$, where $N_1$ is the number of
particles per grid cell. Since the gas is approximately isodense and
isothermal, we can assume that the friction time is independent of the
local state of the gas at the position of a particle. We also assume that the
friction time
does not depend on the velocity difference between the particle and the gas.
This is valid in the Epstein regime, but also in the Stokes regime when the
flow around the grains
is laminar \citep{Weidenschilling1977}. For conditions typical for a
protoplanetary disk at a radial distance of 5 AU from the central object, a
given Stokes number $\varOmega_0 \tau_{\rm f}$ corresponds to the grain radius
measured in meters \citep[e.g.][]{Johansen+etal2006}, although this 
depends somewhat on the adopted disk model. We include the friction
force contribution to the computational time-step $\delta t$ by requiring that
the time $(\delta t)_{\rm fric}=\tau_{\rm f}/(1+\epsilon)$ is resolved at least
five times in a time-step. This restriction is strongest for small grains and
for large dust-to-gas ratios, whereas the Courant time-step of the gas
dominates otherwise.
\begin{figure*}
  \includegraphics{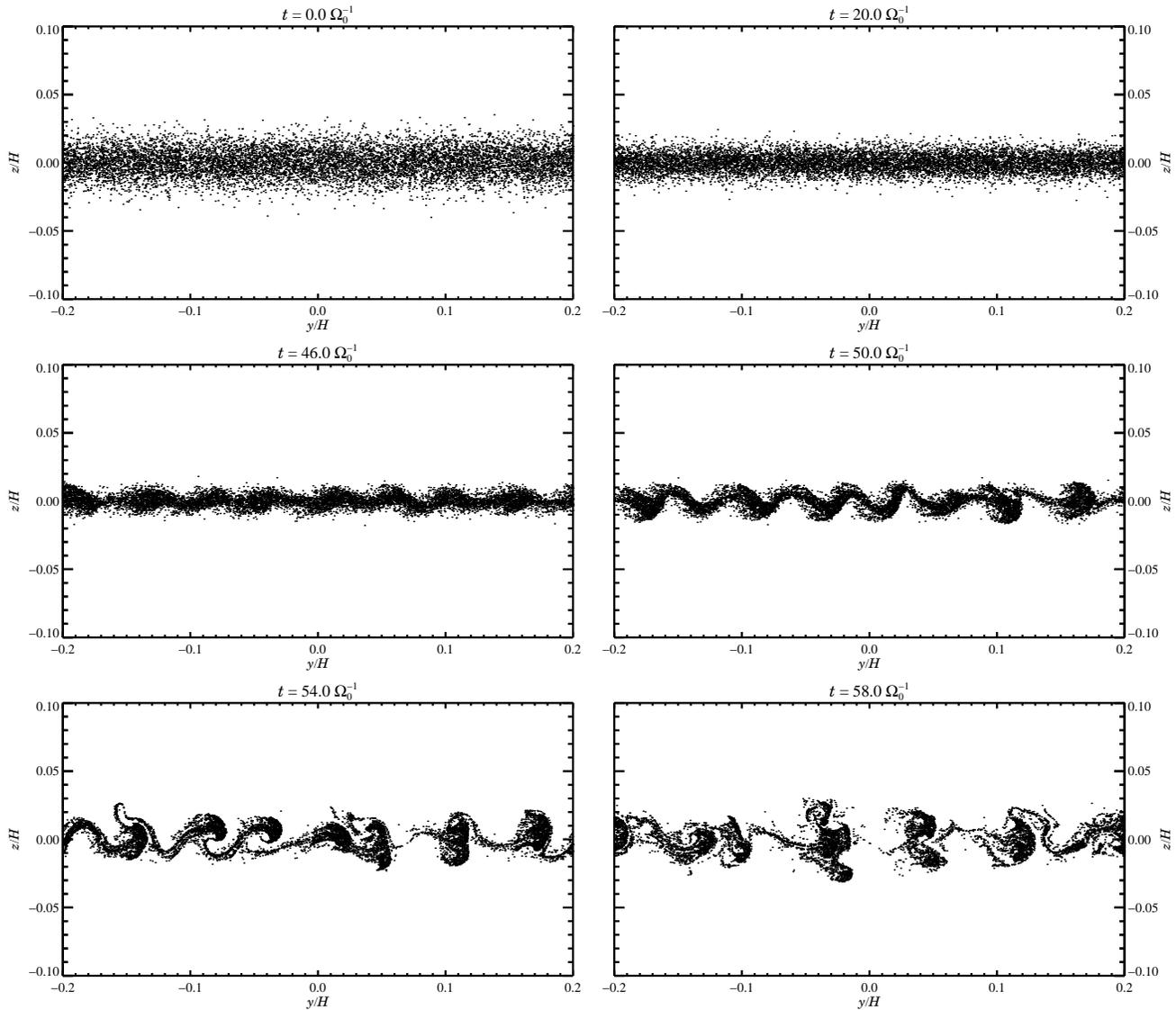}
  \caption{The onset of the Kelvin-Helmholtz instability for 
      cm-sized pebbles with $\varOmega_0 \tau_{\rm f}=0.02$. The initial
      Gaussian particle distribution falls towards the mid-plane of the disk on
      the characteristic time-scale of $t_{\rm grav}=1/(\varOmega_0^2 \tau_{\rm
      f})\approx50\varOmega_0^{-1}$. The increased vertical shear in the gas
      rotation velocity eventually makes the disk unstable to the KHI, forming
      waves that finally break as the turbulence goes into its non-linear
      state.}
  \label{f:particles_panel}
\end{figure*}

The particle positions change due to the velocity of the particles as
\begin{eqnarray}
  \frac{\de x^{(i)}}{\de t} &=& 0 \, , \\
  \frac{\de y^{(i)}}{\de t} &=& v_y^{(i)} \, , \\
  \frac{\de z^{(i)}}{\de t} &=& v_z^{(i)} \, .
\end{eqnarray}
Because the simulations are done in two dimensions, we have not allowed
particles to move in the $x$-direction. The particles are still allowed to have
a radial velocity component. This is equivalent to assuming that all radial
derivatives are zero, so that no advective transport occurs in this direction.

\section{RICHARDSON NUMBER}\label{ch:richardson}

Before we discuss the setup of the numerical simulations and the results, we
describe in this section some of the analytical results that are already known
about the KHI.

The stability of a shear flow can be characterized through the Richardson
number ${\rm Ri}$, defined as
\begin{equation}
  {\rm Ri} = \frac{g_z \dpa \ln (\rho+\rho_{\rm d})/\dpa z}{(\dpa u_y/\dpa z)^2} \, .
\end{equation}
The Richardson number quantifies the fact that vertical gravity $g_z$ and
density stratification of gas and dust $\dpa \ln (\rho+\rho_{\rm d})/\dpa z$ are stabilizing effects, whereas
the shear $\dpa u_y/\dpa z$ is destabilizing. As shown by Chandrasekhar from
very simple considerations of the free energy that is present in a stratified
shear flow, a flow with ${\rm Ri}>1/4$ is always stable, whereas ${\rm
Ri}<1/4$ is necessary, but not sufficient, for an instability \citep[][p.
491]{Chandrasekhar1961}. These derivations do, however, not include the effect
of the Coriolis force, a point that we shall return to later.

For dust-induced shear flows in protoplanetary disks, S98 derived an expression
for the vertical density distribution of dust in a protoplanetary disk that is
marginally stable to the KHI, i.e.\ where the gas flow has a constant
Richardson number equal to the critical Richardson number for stability ${\rm
Ri}_{\rm c}$. For small dust grains the dust-to-gas ratio $\epsilon(z)$ in this
state can be written as
\begin{equation}\label{eq:epsz_Riconst}
  \epsilon(z) =
  \left\{
    \begin{array}{cl}
      \frac{1}{\sqrt{z^2/H_{\rm d}^2 + 1/(1+\epsilon_1)^2}} - 1 & {\rm for}
      \,\,|z|<z_{\rm d} \\
      0 & {\rm for}\,\,|z|\ge z_{\rm d}
    \end{array}
  \right. \, ,
\end{equation}
where $\epsilon_1$ is the dust-to-gas ratio in the mid-plane, $z_{\rm d}=H_{\rm
d}\sqrt{1-1/(1+\epsilon_1)^2}$, and $H_{\rm d}$ is
the width of the dust layer. The effect of self-gravity between the dust grains
has been ignored. The width of the dust layer can furthermore be written as
\begin{equation}\label{eq:Hd}
  H_{\rm d} = \sqrt{{\rm Ri}_{\rm c}} \frac{|\beta|}{2 \gamma} H \, ,
\end{equation}
where $\beta$ is the radial pressure gradient parameter introduced in
\Sec{ch:gas_equations}, $\gamma$ is the ratio of specific heats and $H$ is the
scale-height of the gas.  For ${\rm Ri}_{\rm c}=1/4$ and $\gamma=5/3$
\begin{equation}
  H_{\rm d}/H = \frac{3}{20} |\beta| \, ,
\end{equation}
so the width of the marginally stable dust layer is a few percent of a gas
scale height.

The expression in \Eq{eq:epsz_Riconst} allows for two types of dust
stratification in the marginally stable disk. For $\rho_{\rm d}\ll\rho$ the
dust density is constant around the mid-plane, whereas for $\rho_{\rm
d}\gg\rho$, a cusp of very high dust density can exist around the mid-plane
(S98).  Such a cusp can form for two reasons when the dust-to-gas ratio is
above unity.  Firstly because the gas flow is forced to be Keplerian in such a
large region around the mid-plane that the vertical shear is reduced there,
stabilizing against the KHI, and secondly because it requires a lot of energy
to lift up so much dust away from the mid-plane. This effect has been
interpreted by \cite{YoudinShu2002} as the gas only being able to lift up its
own equivalent mass due to KHI.

\section{INITIAL CONDITION}\label{ch:initial_condition}

The initial condition of the gas is an isothermal, stratified disk with a scale
height $H$. The density depends on the height over the mid-plane $z$ as 
\begin{equation}
  \rho(z) = \rho_1 \ee^{-z^2/(2 H^2)} \, ,
\end{equation}
where $\rho_1$ is the density in the mid-plane.  The scale height is $H=c_{\rm
s}/\varOmega_0$, where $c_{\rm s}$ determines the constant initial temperature,
to sustain hydrostatic equilibrium in the vertical direction. From the
definition of the gas column density $\varSigma$, we can calculate the
mid-plane density as $\rho_1=\varSigma/(\sqrt{2\pi} H)$.  There is no similar
equilibrium to set the dust density in the disk. Thus we distribute the
particles in a Gaussian way around the mid-plane with a scale height $H_{\rm
d}$, a free parameter, and normalize the distribution so that $\varSigma_{\rm
d}=\epsilon_0 \varSigma$, where $\epsilon_0$ is the global dust-to-gas ratio in
the disk.

The constant global pressure gradient force effectively decreases the
radial gravity
felt by the gas, and thus the orbital speed is no longer Keplerian, but
slightly smaller. The sub-Keplerian velocity $u_y^{(0)}$ is given by
\begin{eqnarray}
  u_y^{(0)} = \frac{\beta}{2\gamma} c_{\rm s} \, .
\end{eqnarray}
This expression is obtained by setting $u_x = \dpa u_x/\dpa t = \epsilon = 0$
in \Eq{eq:eqmot1}. The dust grains, on the other hand, do not feel the global
pressure gradient and would thus in the absence of friction move on a Keplerian
orbit with $v_y^{(i)}=0$. The drag force from the gas, however, forces the dust
grains to move at a speed that is below the Keplerian value, at least when the
dust-to-gas ratio is low. When the dust-to-gas ratio approaches unity or even
larger, the gas is forced by the dust to move with Keplerian speed.  The
equilibrium gas and dust velocity  can be calculated as a function of
dust-to-gas ratio \citep{Nakagawa+etal1986}, but we choose to simply start the
gas with a sub-Keplerian velocity and the dust with a Keplerian velocity, and
then let the velocities approach the equilibrium dynamically (this happens
within a few friction times). This way we have checked that the numerical
solution indeed approaches the expressions by \cite{Nakagawa+etal1986} for all
the velocity components of the gas and the dust, which serves as a control that
the friction force from the dust particles on the gas has been correctly implemented in
the code.

In the equilibrium state the gas has a positive radial velocity in the
mid-plane, but this does not lead to any change of the gas density, since we
have ignored the advection of the global density in the continuity equation.
The effect of an outwards-moving mid-plane on the global dynamics of a
protoplanetary disk is a promising topic of future research, but it is beyond
the scope of this paper.

Periodic boundary conditions are used for all variables in the azimuthal
direction. In the vertical direction we impose a condition of zero vertical
velocity, whereas the two other velocity components have a zero derivative
condition over the boundary. The logarithmic mass density and the entropy have
a condition of vanishing third derivatives over the vertical boundary.

We run simulations for three different grain sizes, $\varOmega_0 \tau_{\rm
f}=0.02,0.1,1.0$, respectively. When assuming compact spherical grains, these
numbers correspond to sizes of centimeters (pebbles), decimeters (rocks), and
meters (boulders), respectively.  The computation parameters are listed in
\Tab{t:parameters}. The size of the box is set according to the vertical extent
of the dust layer in the state of self-sustained Kelvin-Helmholtz turbulence.
We make sure that the full width of the dust layer fits at least five times
vertically in the box to avoid any effect of the vertical boundaries on the
mid-plane. The azimuthal extent of the box is set so that the full width of the
dust layer fits at least ten times in this direction. Thus the unstable modes
of the Kelvin-Helmholtz turbulence, which have a wavelength that is similar to
the width of the dust layer, are very well-resolved.

Since the ratio of particles to grid points is $N_1\approx12$, the computation
time is strongly dominated by the particles. Each particle needs to ``know''
the gas velocity field at its own position in space, to calculate the friction
forces.  For parallel runs we distribute the particles among the different
processors so that the position of each particle is within its ``host''
processor's space interval. As we show in \Sec{ch:results}, the particles tend
to have a strongly non-axisymmetric density distribution in the
Kelvin-Helmholtz turbulence. This clumping means that the number of particles
on the individual processors varies by a factor of around five, thus slowing
the code down by a similar factor compared to a run where the particles were
equally distributed over the processors. For this reason we used 32 Opteron
processors, each with 2.2 GHz CPU speed and Infiniband interconnections, for
around one week for each run.
\begin{figure}
  \includegraphics{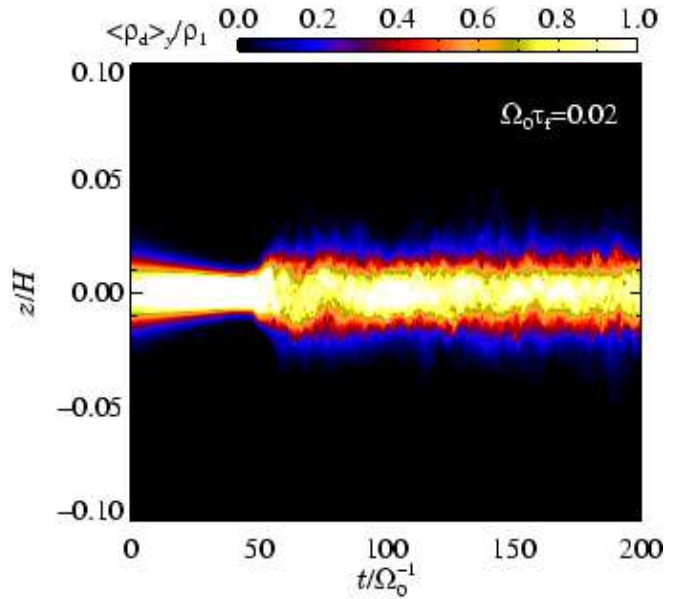}
  \caption{Contour plot of the dust density of cm-sized pebbles averaged over
      the azimuthal $y$-direction, as a function of time $t$ and height over
      the mid-plane $z$.  After the Kelvin-Helmholtz instability sets in and
      saturates into turbulence, the width of the dust layer stays
      approximately constant. The black regions contain no particles at all.}
  \label{f:rhodmz_zt_tauf0.02}
\end{figure}
\begin{figure}
  \includegraphics{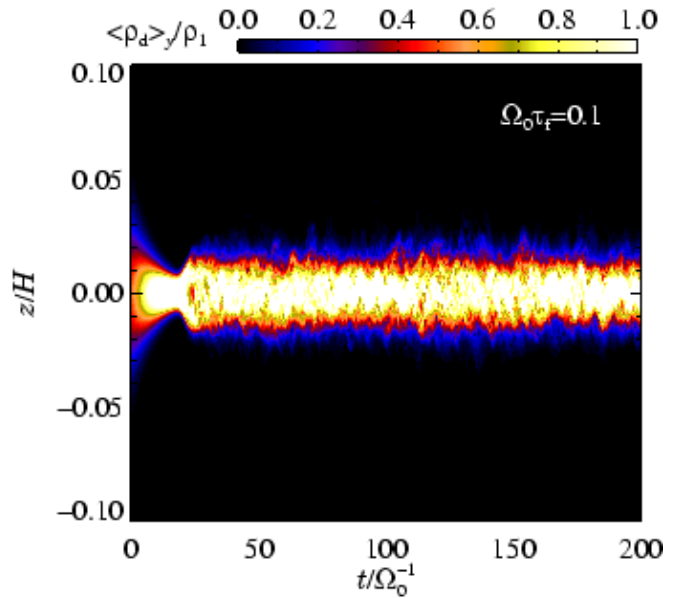}
  \caption{Same as \Fig{f:rhodmz_zt_tauf0.02}, but for dm-sized rocks with
      $\varOmega_0 \tau_{\rm f}=0.1$. The sedimentation time-scale is much
      faster than in \Fig{f:rhodmz_zt_tauf0.02}, but the width of the dust
      layer in the self-sustained state of turbulence is approximately the
      same.}
  \label{f:rhodmz_zt_tauf0.1}
\end{figure}
\begin{figure}
  \includegraphics{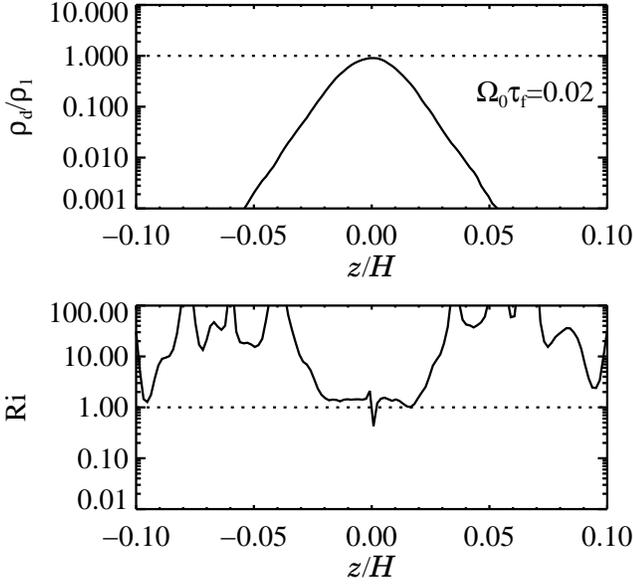}
  \caption{Dust density and Richardson number of grains with
      $\varOmega_0 \tau_{\rm f}=0.02$ averaged over the azimuthal
      direction and over time. The dust-to-gas ratio in the mid-plane is close
      to unity and falls rapidly outwards. The Richardson number is
      approximately constant in the mid-plane and has a value around unity.}
  \label{f:Rimt_z_t0.02}
\end{figure}
\begin{figure}
  \includegraphics{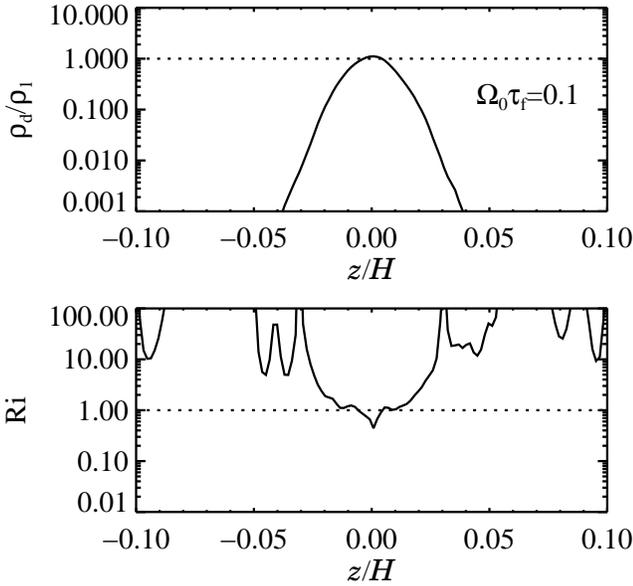}
  \caption{Same as in \Fig{f:Rimt_z_t0.02}, but for $\varOmega_0 \tau_{\rm
      f}=0.1$. Although the density in the mid-plane is similar to the value
      for smaller grains, the dust is more settled and has less pronounced
      wings away from the mid-plane. The Richardson number is again around
      unity in the mid-plane.}
  \label{f:Rimt_z_t0.1}
\end{figure}
\begin{figure}
  \includegraphics{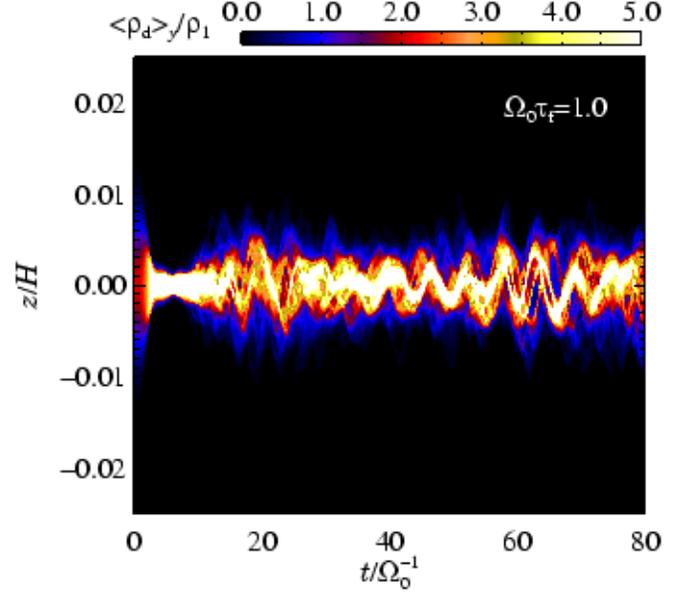}
  \caption{Same as \Fig{f:rhodmz_zt_tauf0.1}, but for m-sized boulders with
      $\varOmega_0 \tau_{\rm f}=1.0$. The equilibrium scale height of the dust
      is around 10 times lower than for the smaller grains.}
  \label{f:rhodmz_zt_tauf1.0}
\end{figure}
\begin{figure}
  \includegraphics{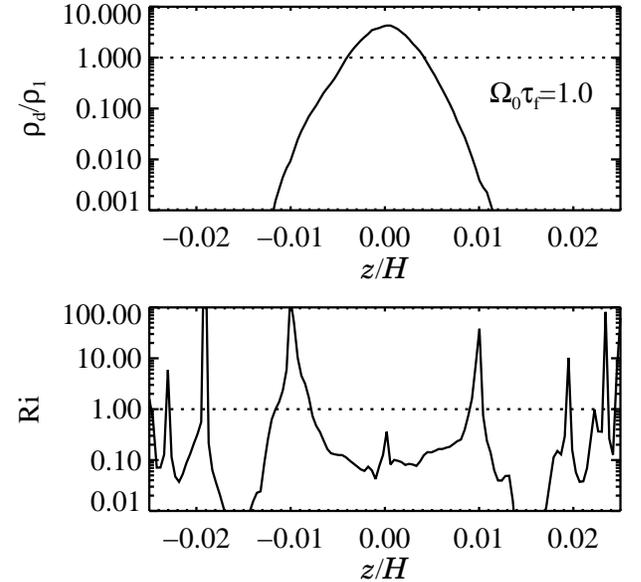}
  \caption{Same as in \Fig{f:Rimt_z_t0.02}, but for $\varOmega_0 \tau_{\rm
      f}=1.0$. The Richardson number around the mid-plane is here way lower
      than for the smaller grains, around 0.1. This is caused by the extremely
      rapid settling of m-sized boulders to the mid-plane.}
  \label{f:Rimt_z_t1.0}
\end{figure}

\section{DYNAMICS AND DENSITY OF SOLIDS}\label{ch:results}

\begin{figure*}
  \includegraphics{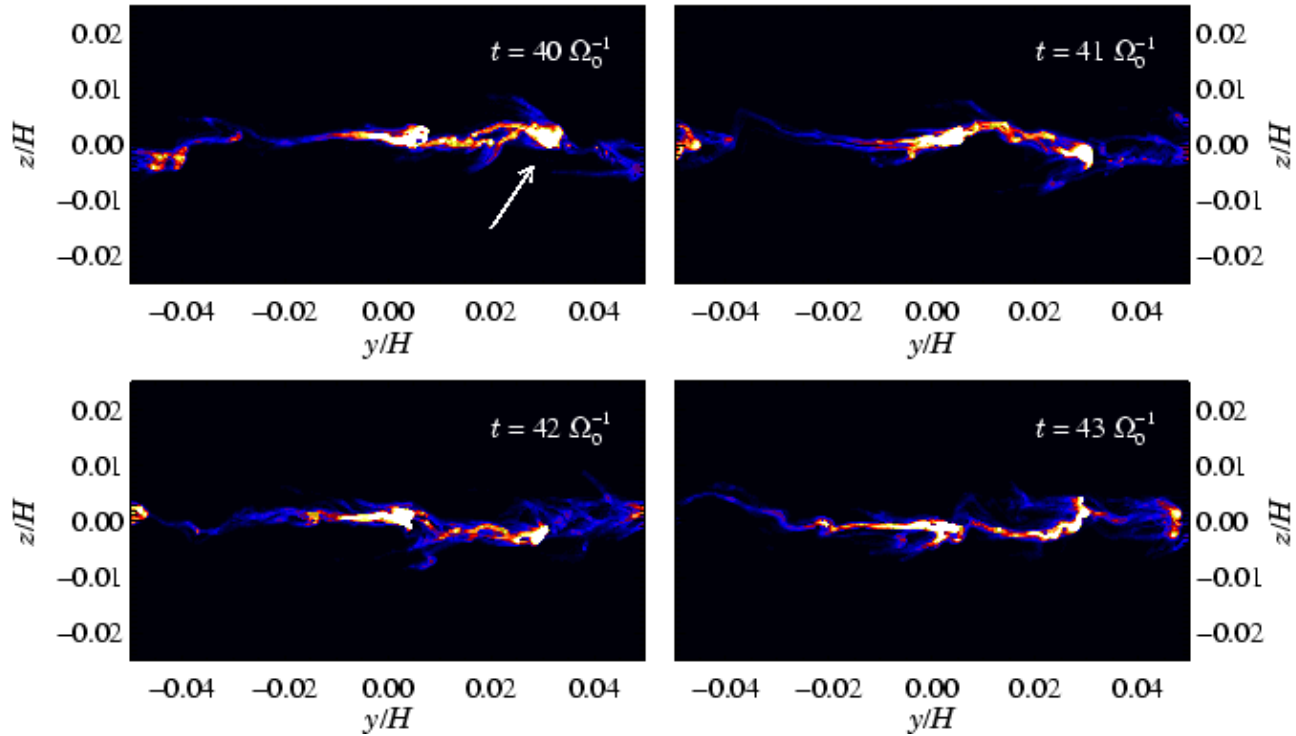}
  \caption{Contour plot of the particle density for $\varOmega_0 \tau_{\rm
      f}=1.0$. A clump, indicated by the arrow, oscillates around the
      mid-plane, a type of motion that is only allowed for particles with
      $\varOmega_0 \tau_{\rm f}>0.5$.}
  \label{f:particles_panel2}
\end{figure*}
In this section we focus on the dynamics and the density of the dust particles.
The linear growth rate of the Kelvin-Helmholtz instability and the statistical
properties of the Kelvin-Helmholtz turbulence are treated in the next two
sections.

Some representative snapshots of the particle positions for run A ($\varOmega_0
\tau_{\rm f}=0.02$, or cm-sized pebbles) are shown in \Fig{f:particles_panel}.
The particles, with an initial Gaussian density distribution, settle to the
mid-plane due to gravity, on the characteristic time-scale $t_{\rm
grav}=1/(\tau_{\rm f}\varOmega_0^2)\approx50$.  When the width of the layer has
decreased to around $0.01$ scale heights (the two middle panels of
\Fig{f:particles_panel}), some wave pattern can already be seen in the dust
density. It is the most unstable $u_z(y)$ mode, with a wavelength that is
comparable to the vertical width of the layer, that is growing in amplitude.
Some shear times later, in the two bottom panels of \Fig{f:particles_panel},
the growing mode enters the non-linear regime, and impressive patterns of
breaking waves appear. The simulation then goes into a state of fully developed
Kelvin-Helmholtz turbulence.

\subsection{Pebbles and Rocks}

In \Figs{f:rhodmz_zt_tauf0.02}{f:rhodmz_zt_tauf0.1} we show azimuthally
averaged dust density contours as a function of time for grains with friction
time $\varOmega_0 \tau_{\rm f}=0.02$ (run A) and $\varOmega_0 \tau_{\rm f}=0.1$
(run B, decimeter-sized rocks), respectively. These two grain sizes show very
similar behavior with time. In the beginning the particles move towards the
mid-plane unhindered because of the lack of turbulence. The sedimentation
happens much faster in run B than in run A because of the different grain
sizes. When the KHI eventually sets in, the dust layer is puffed up again and
quickly reaches an equilibrium configuration where the vertical distribution of
dust density is practically unchanged with time.

One can already here suspect that the equilibrium dust density is indeed, as
predicted analytically by S98, distributed in such a way that the flow has a
constant Richardson number.  In \Figs{f:Rimt_z_t0.02}{f:Rimt_z_t0.1} we plot
the time-averaged dust density and the Richardson number as a function of
vertical height over the mid-plane, again for the two small grain sizes. The
dust-to-gas ratio reaches unity in the mid-plane and drops down rapidly away
from the mid-plane.  For run A the Richardson number is approximately constant,
just above unity, in the region around the mid-plane that has a significant
dust density. For run B the value of the Richardson number is also constant,
although somewhat smaller than for the centimeter-sized grains, because the
more rapid sedimentation of these larger grains allows the disk to sustain a
stronger vertical shear.

\subsection{Boulders}

For bodies with $\varOmega_0 \tau_{\rm f}=1.0$ (run C, m-sized boulders), the
azimuthally averaged dust density is shown in \Fig{f:rhodmz_zt_tauf1.0}. These
meter-sized boulders fall rapidly to the mid-plane, on a time-scale of one
shear time, because they are not as coupled to the gas as smaller grains. The
scale height of the boulders is very small, less than one percent of the scale
height of the gas, because the grains are falling so fast that the disk can
sustain a much lower Richardson number than the critical.  This is also evident
from \Fig{f:Rimt_z_t1.0}. The Richardson number is well below unity, around
$0.1$, where significant amounts of dust is present.

A major difference between the large grains and the small grains is the
presence of bands in \Fig{f:rhodmz_zt_tauf1.0}. The dust grains are no longer
smoothly distributed over $z$, but rather appear as clumps that oscillate
around the mid-plane.  The oscillation of a single clump is evident from
\Fig{f:particles_panel2}.  Here the dust density contours are plotted at four
times separated by one shear time. The clump indicated by the arrow is
oscillating around the mid-plane.  Such oscillatory behavior is also be
expected from the following considerations. Friction ensures that small grains
arrive at the mid-plane with zero residual velocity, whereas larger grains
perform damped oscillations around $z=0$ with a damping time of one friction
time. The distinction between the two size regimes can be derived by looking at
the differential equation governing vertical settling of particles,
\begin{equation}
  \frac{\de v_z(t)}{\de t} =
      -\varOmega_0^2 z - \frac{1}{\tau_{\rm f}} v_z \, .
\end{equation}
This second order, linear ordinary differential equation can be solved
trivially \citep[e.g.][]{Nakagawa+etal1986}. The result is a split between two
types of solutions, depending on the value of $\varOmega_0 \tau_{\rm f}$. For
$\varOmega_0 \tau_{\rm f}\leq0.5$ the solution is a purely exponentially
decaying function. On the other hand, for $\varOmega_0 \tau_{\rm f}>0.5$ the
solutions are damped oscillations with a characteristic damping time of around
one friction time. For a friction time around unity, the amplitude of the dust
density in a laminar disk would still become virtually zero in just a few
friction times. This is not the case in \Fig{f:rhodmz_zt_tauf1.0} where the
dust scale height stays approximately constant for at least 80 friction times
(the end of the simulation). The KHI is continuously pumping energy into the
dust layer at the same rate as the oscillations are damped.

\begin{figure}
  \includegraphics{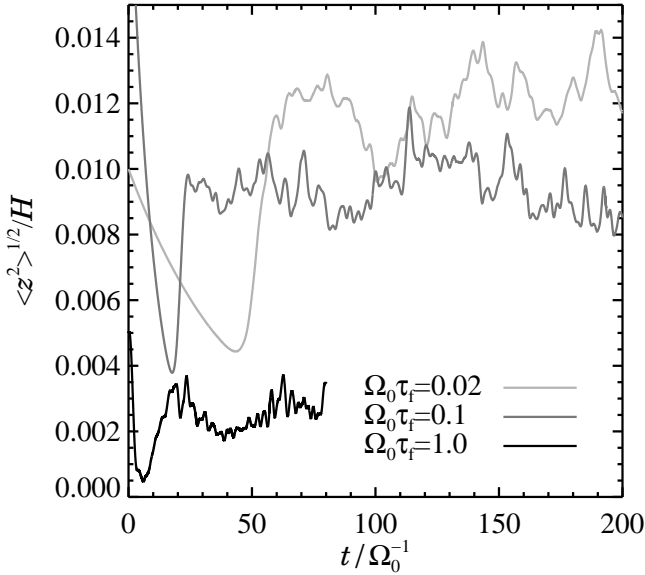}
  \caption{The root-mean-square $z$-coordinate of the particles for the
      different runs. For $\varOmega_0 \tau_{\rm f}=0.02$ and $\varOmega_0
      \tau_{\rm f} = 0.1$, the width of the dust layer is around 1\% of a gas
      scale height, whereas for meter-sized boulders with $\varOmega_0
      \tau_{\rm f}=1.0$, the strong sedimentation results in a much lower
      width, only around 0.25\% of a scale height.}
  \label{f:zpm_t}
\end{figure}
In \Fig{f:zpm_t} we plot the root-mean-square value of the $z$-coordinate of
the particles as a function of time for all three runs. The calculated ``scale
height'' of the cm-sized and dm-sized particles are very similar, although the
larger particles have a bit lower scale height than the smaller particles. The
m-sized particles have a scale height of about one quarter of a percent of that
of the gas.

\subsection{Maximum Density and Clumping}

It is of great relevance for planetesimal formation to find the highest dust
density that is permitted in the saturated Kelvin-Helmholz turbulence. Both
coagulation and gravitational fragmentation depend strongly on the mass density
of the dust layer. High densities can occur not only when the gas flow or the
size of the boulders allow for a high mid-plane density, but also in certain
points of the turbulent flow where dust particles tend to accumulate. The
latter can only be explored in computer simulations, so we examine the maximum
dust density in any grid point in more detail in this section.

\begin{figure}
  \includegraphics{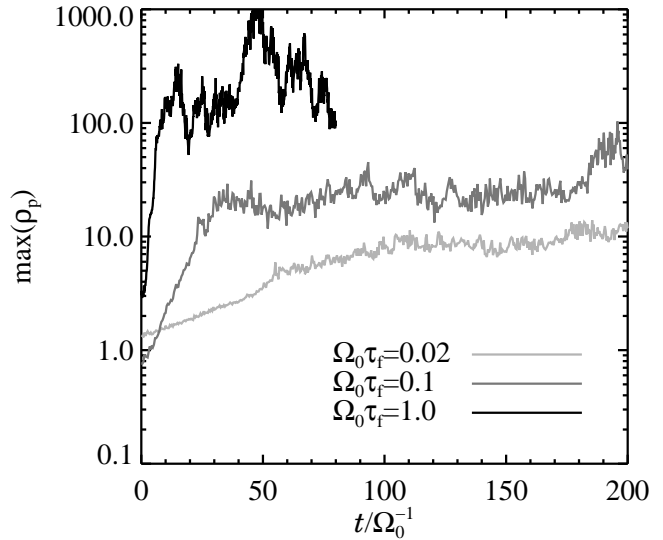}
  \caption{The maximum dust density in any grid cell as a function of time, in
      units of the gas mid-plane density, as a function of time. The value is
      much higher than the azimuthally averaged mid-plane densities (presented
      in previous figures).}
  \label{f:rhopmax_t}
\end{figure}
The maximum mass density of dust particles in any grid cell is plotted in
\Fig{f:rhopmax_t} as a function of time. Even though the mid-plane dust-to-gas
ratio is on the average of the order unity for all grain sizes, the maximum
density is much higher at all times, especially for meter-sized particles where
the maximum dust-to-gas ratio can be up to one thousand. Decimeter-sized
particles have a maximum dust-to-gas ratio of around 20 at all times, whereas
the value for centimeter-sized particles is around 10. This is potentially
important for building planetesimals. Even if the critical density for
gravity-aided planetesimal formation is not reached globally, this is still
possible in certain regions of the turbulent flow. Such a gravoturbulent
formation of planetesimals was proposed by \cite{Johansen+etal2006} to lead to
the formation of planetesimals in a magnetorotationally turbulent gas.

\begin{figure}
  \includegraphics{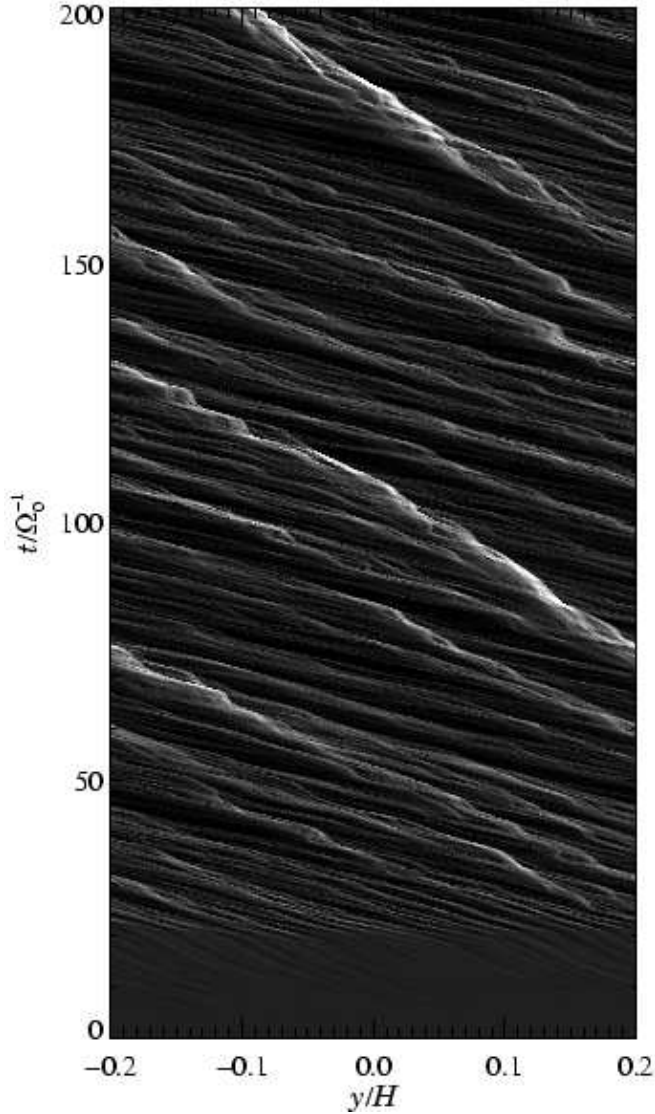}
  \caption{Vertically averaged dust density of rocks with
      $\varOmega_0 \tau_{\rm f}=0.1$ as a function of azimuthal coordinate $y$
      and time $t$. The clumping mechanism is evident from the plot. Regions of
      high dust-to-gas ratio (light) move slower than regions of low
      dust-to-gas ratio (gray), seen in the different slopes of the bright and
      dark wisps on the plot, causing the high density clumps to be
      continuously fed by low density material. One also sees the rarefaction
      tail going to the left of the dense clumps and the shock front that is
      formed against the sub-Keplerian stream.}
  \label{f:rhodmy_yt}
\end{figure}
\begin{figure}
  \includegraphics{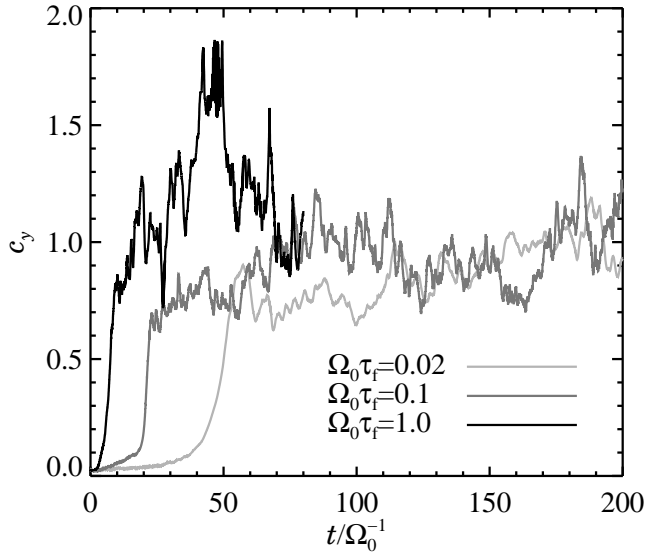}
  \caption{Azimuthal clumping factor $c_y$ versus time for all three grain
      sizes. A value of around unity corresponds to strong clumping with the
      average point being $100\%$ away in density from the average density.}
  \label{f:clumpingy_t}
\end{figure}

In \Fig{f:rhodmy_yt} we plot contours of the vertically averaged dust density
as a function of azimuthal coordinate $y$ and time $t$ for decimeter-sized
rocks (run B).  It is evident that the dust density has a strong
non-axisymmetric component once the Kelvin-Helmholtz turbulence is fully
developed. Dense clumps are seen as white stripes, while regions of lower
dust-to-gas ratio are gray.  A simple way to quantify the amount of
non-axisymmetry is to look at the mean deviation of the dust density from the
average density.  We define the {\it azimuthal clumping factor} $c_y$ as
\begin{equation}
  c_y = \frac{\sqrt{\langle[n_y(y)-\langle n_y(y) \rangle]^2\rangle}}{\langle
  n_y(y) \rangle} \, .
\end{equation}
Here $n_y(y) \equiv \langle n(y,z) \rangle_z$ is the dust number density
averaged over the vertical direction. Axisymmetry implies $c_y=0$, whereas
higher values of $c_y$ imply stronger and stronger non-axisymmetry. We plot in
\Fig{f:clumpingy_t} the azimuthal clumping factor as a function of time for all
three values of the friction time. For centimeter and decimeter particles the
clumping is of the order unity, or in other words, the average grid point has a
density variation from the average that is on the same order as the average,
i.e.\ very strong clumping. For meter-sized particles the azimuthal clumping is
even stronger.

The tendency for the dust particles to clump is a consequence of the
sub-Keplerian velocity of the gas. Turning again to \Fig{f:rhodmy_yt}, one sees
that brighter regions move at a lower speed (relative to the Keplerian speed)
than dark regions do. The speed of a clump is evident from the absolute value
of the angle between the tilted time-space wisp and the time-axis. Bright wisps
have a higher angle with the time-axis than dark wisps. This instability is
very related to the {\it streaming instability} found by
\cite{YoudinGoodman2005}, although in our simulations the clumping happens in
the $(y,z)$-plane and not the $(x,z)$-plane as in the analysis by Youdin \&
Goodman.
Still the instability is powered in both planes by the dependence of the dust
velocity on the dust-to-gas ratio, so we consider the instability in the
$(y,z)$-plane a special case of the streaming instability. We refer to
\cite{YoudinGoodman2005} for a linear stability analysis of the coupled gas and
dust flow. In the rest of this section we instead focus on describing the
non-linear outcome of the streaming instability with a simple analogy to a
hydrodynamical shock.

Using the derivations given by \cite{Nakagawa+etal1986} for the equilibrium gas
and dust velocities as a function of the dust-to-gas mass ratio $\epsilon$, one
can write up the azimuthal velocity component of the dust as
\begin{equation}\label{eq:wy_equi}
  w_y = \frac{1+\epsilon}{(1+\epsilon)^2+(\varOmega_0 \tau_{\rm f})^2} u_y^{(0)}
  \, .
\end{equation}
Thus clumps with a high dust-to-gas ratio move slower, relative to the
Keplerian speed, than clumps with a low dust-to-gas ratio. The small clumps
crash into the big clumps and form larger structures. At the same time, the
large clumps steepen up against the direction of the sub-Keplerian flow and
develop an escaping tail downstream. This is qualitatively similar to a shock.
Considering the continuity equation of the dust-to-gas ratio
\begin{equation}\label{eq:eps_const}
  \frac{\dpa \epsilon}{\dpa t} = -w_y \frac{\dpa \epsilon}{\dpa y} -
  \epsilon\frac{\dpa w_y}{\dpa y} \, ,
\end{equation}
and inserting \Eq{eq:wy_equi} in the limit of small Stokes numbers
$\varOmega_0\tau_{\rm f}\ll1$, \Eq{eq:eps_const} can be reduced to
\begin{equation}
  \frac{\dpa \epsilon}{\dpa t} =
      -\frac{u_y^{(0)}}{(1+\epsilon)^2} \frac{\dpa \epsilon}{\dpa y} \, ,
\end{equation}
qualitatively similar to the advection equation of fluid dynamics. The shock
behavior of the clumps arises because the advection velocity
$u_y^{(0)}/(1+\epsilon)^2$ depends on the dust-to-gas ratio.

\subsection{Varying the Global Dust-to-gas Ratio}

It is of great interest to investigate the dependence of the mid-plane dust
density on the global dust-to-gas ratio in the saturated state of
Kelvin-Helmholtz turbulence. Increasing the dust-to-gas ratio beyond the
interstellar value should potentially lead to the creation of a very dense
mid-plane of dust that the gas is not able to lift up, making the dust layer
dense enough to undergo gravitational fragmentation
\citep{Sekiya1998,YoudinShu2002,YoudinChiang2004}.

The analytically predicted mid-plane dust-to-gas ratio $\epsilon_1$ is found by
applying the normalization
\begin{equation}\label{eq:eps_norm}
  \int_{-\infty}^\infty \rho_1 \epsilon(z) \de z = \varSigma_{\rm d}
\end{equation}
to the constant Richardson number dust density of \Eq{eq:epsz_Riconst}.  Here
$\varSigma_{\rm d}$ is the dust column density, a free parameter that we set
through the global dust-to-gas ratio $\epsilon_0$ as $\varSigma_{\rm
d}=\epsilon_0 \varSigma$. We have approximated the gas density by the gas
density in the mid-plane $\rho_1$, because for $z\ll H$, the variation in gas
density is insignificant compared to the variation in dust density. We proceed
by inserting the expression for the dust density in a disk with a constant
Richardson number, from \Eq{eq:epsz_Riconst}, into the integral in
\Eq{eq:eps_norm}. Defining the parameter
\begin{equation}\label{eq:chi}
  \chi = \frac{\sqrt{\epsilon_1 (2+\epsilon_1)}}{1+\epsilon_1} \, ,
\end{equation}
the integration yields
\begin{equation}\label{eq:chi2}
  -2 \chi + \ln\left(\frac{1+\chi}{1-\chi}\right) =
      \frac{\varSigma_{\rm d}}{H_{\rm d} \rho_1} \, .
\end{equation}
This is a transcendental equation that we solve numerically for $\chi$ as a
function of the input parameters $H_{\rm d}$, given by \Eq{eq:Hd}, and
$\varSigma_{\rm d}$. Once $\chi$ is calculated, then the dust-to-gas ratio in
the mid-plane $\epsilon_1$ is known from \Eq{eq:chi}.

\begin{figure}
  \includegraphics{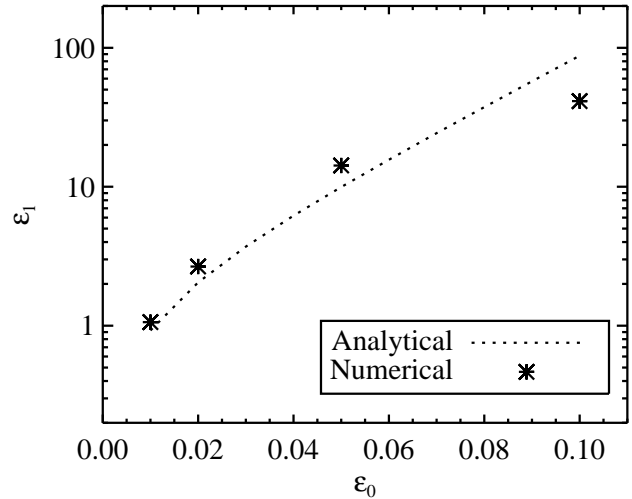}
  \caption{Mid-plane dust-to-gas ratio $\epsilon_1$ as a function of global
      dust-to-gas ratio $\epsilon_0$. The dotted line shows the analytical
      value for a dust density with a constant Richardson number of ${\rm
      Ri}_{\rm c}=1.0$, while the stars show the result of the numerical
      simulations for different values of the global dust-to-gas ratio. The
      results agree nicely, giving support to the idea that a dust-to-gas ratio
      that is higher than the interstellar value can give rise to high enough
      mid-plane dust density for a gravitational instability in the dust layer.}
  \label{f:eps1_eps0}
\end{figure}
In \Fig{f:eps1_eps0} we plot the analytical mid-plane dust-to-gas ratio
$\epsilon_1$ as a function of the global dust-to-gas ratio $\epsilon_0$ (dotted
line). The non-linear behavior of $\epsilon_1$ is evident, and already for
$\epsilon_0=0.1$ does the mid-plane dust-to-gas ratio approach
$\epsilon_1=100$, which should be enough to have a gravitational instability in
the dust layer. We also run numerical simulations with an increased global
dust-to-gas ratio (runs Be2, Be5 and Be10, see \Tab{t:parameters}) to see if a
mid-plane dust density cusp develops as predicted. The resulting mid-plane
dust-to-gas ratio is indicated with stars in \Fig{f:eps1_eps0}. To avoid having
a very low time-step, because of the strong friction that the dust exerts on
the gas when the global dust-to-gas ratio is increased, we have locally
increased the friction time in regions of very high dust-to-gas ratio. This
approach conserves total momentum because the friction force on the gas and on
the dust are made lower at the same time. In the regions where the friction
time is increased, the dust-to-gas ratio is so high that gas is dragged along
with the particles anyway, so the precise value of the friction time does not
matter.  As seen in \Fig{f:eps1_eps0} the mid-plane dust-to-gas ratio does
indeed increase non-linearly with global dust-to-gas ratio, following a curve
that is within a factor of two of the analytical curve. This gives support to
the theory that an increased global dust-to-gas ratio, e.g.\ due to solids that
are transported from the outer part of the disk into the inner part, can lead
to such a high dust-to-gas ratio in the disk mid-plane that the dust layer
fragments to form planetesimals \citep{YoudinShu2002}.

\section{GROWTH RATES}\label{ch:growth_rates}

The simulations presented so far all imply a critical Richardson number that is
of the order unity, rather than the classically adopted value of $1/4$. This
confirms the findings of GO05 that shear flows with a Richardson number that is
higher than the classical value are actually unstable when the Coriolis force
is included in the calculations. To quantify the linear growth of the
instability we have run simulations of initial conditions with a constant
Richardson number and measured the growth rates of the KHI.  Because we are
only interested in the linear regime, we have chosen for simplicity to treat
dust as a fluid rather than as particles. Thus we solve equations similar to
\Eqss{eq:eqmotp1}{eq:eqmotp3} for the dust velocity field $\vc{w}$ including an
advection term $(\vc{w}\cdot\nab)\vc{w}$. A continuity equation similar to
\Eq{eq:conteq} for the logarithmic dust number density $\ln n$ is solved at the
same time.

\begin{figure}
  \includegraphics{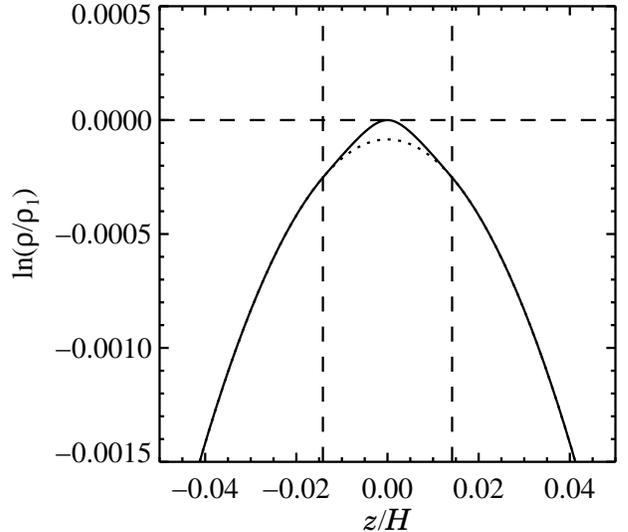}
  \caption{The logarithmic gas density as a function of height over the
      mid-plane $z$ in the presence of falling dust.  The drag force exerted on
      the gas by the falling dust mimics an extra gravity near the mid-plane,
      making the gas scale height slightly lower close to the mid-plane. The
      result is the formation of a cusp, although of a very moderate amplitude
      of about 1/10000 compared to a disk with no dust sedimentation (dotted
      line).}
  \label{f:cusp}
\end{figure}

We consider initial conditions with a constant Richardson number, in the range
between $0.1$ and $2.0$, as given by \Eq{eq:epsz_Riconst}. The dust-to-gas
ratio in the mid-plane $\epsilon_1$ is known from \Eqs{eq:chi}{eq:chi2}.  To
avoid any effects of dust settling, we set the friction time to $\varOmega_0
\tau_{\rm f}=0.001$. The gravitational settling time is then as high as $1000
\varOmega_0^{-1}$, and since this is much longer than the duration of the
linear growth, the effect on the measured growth rate is insignificant. The
vertical velocity of the dust is set to the terminal settling velocity
$w_z=-\tau_{\rm f} \varOmega_0^2 z$. Since the dust velocity is not zero, the
gas will feel some friction from the falling dust. The vertical component of
the equation of motion of the gas is
\begin{equation}\label{eq:eqmot3lin}
  \frac{\dpa u_z}{\dpa t} =
      - \varOmega_0^2 z
      - \frac{1}{\gamma} c_{\rm s}^2 \frac{\dpa \ln \rho}{\dpa z}
      - \frac{\epsilon}{\tau_{\rm f}} (u_z-w_z) \, .
\end{equation}
We insert the terminal velocity expression for the dust velocity into
\Eq{eq:eqmot3lin} and look for equilibrium solutions with $u_z=\dpa u_z/\dpa
t=0$. The equation is then reduced to
\begin{equation}\label{eq:eqmot3lin2}
  0 = -(1+\epsilon) \varOmega_0^2 z
      - \frac{1}{\gamma} c_{\rm s}^2 \frac{\dpa \ln \rho}{\dpa z} \, .
\end{equation}
The drag force exerted by the falling dust on the gas mimics a vertical
gravity, and therefore we have combined it with the regular gravity term. In a
way it {\it is} the gravity on the dust that the gas feels, only it is
transferred to the gas component through the drag force.
One can interpret this as the gas feeling a stronger gravity
$\varOmega_0'=\sqrt{1+\epsilon}\varOmega_0$ in places of high dust-to-gas
ratio, which leads to the creation of a small cusp in the gas density around
the mid-plane. Inserting now \Eq{eq:epsz_Riconst} into \Eq{eq:eqmot3lin2} and
applying the boundary condition $\rho(z=0)=\rho_1$ yields
\begin{equation}\label{eq:lnrho_Riconst}
  \ln \rho(z) =
  \left\{
    \begin{array}{l}
      \ln \rho_1 + \frac{\gamma \varOmega_0^2 H_{\rm d}^2}{c_{\rm s}^2}
          \left[ \frac{1}{1+\epsilon_1} -
          \sqrt{\frac{z^2}{H_{\rm d}^2} + \frac{1}{(1+\epsilon_1)^2}} \right]\\
      \hspace{4cm} {\rm for} \,\,|z|<z_{\rm d}\\
      \ln \rho_1 + \frac{\gamma \varOmega_0^2 H_{\rm d}^2}{c_{\rm s}^2}
          \left[ -\frac{1}{2}\frac{z^2}{H_{\rm d}^2}
          - \frac{\epsilon_1^2}{2(1+\epsilon_1)^2} \right] \\
      \hspace{4cm} {\rm for}\,\,|z|\ge z_{\rm d}
    \end{array}
  \right.
  \, .
\end{equation}
The cusp around the mid-plane, caused by the extra gravity imposed by the
falling dust on the gas, is shown in \Fig{f:cusp}. The variation in density
from a normal isothermal disk is only a few parts in ten thousand, so the
effect is not big. On the other hand, it is important to have a complete
equilibrium solution as the initial condition for the measurement of the linear
growth of the instability, as otherwise dynamical effects could dominate over
the growth.

We measure the linear growth rate of the KHI by prescribing a dust-to-gas ratio
according to \Eq{eq:epsz_Riconst} and a gas density according to
\Eq{eq:lnrho_Riconst}. We then set the velocity fields of gas and dust
according to the expressions derived in \cite{Nakagawa+etal1986}. The width of
a dust layer with a constant Ri depends on the value of Ri according to
\Eq{eq:Hd}, so we have made sure to always resolve the unstable wavelengths by
making the box larger with increasing Ri. The fluid simulations are all done
with a grid resolution of $N_y \times N_z = 256\times128$.

\begin{figure}
  \includegraphics{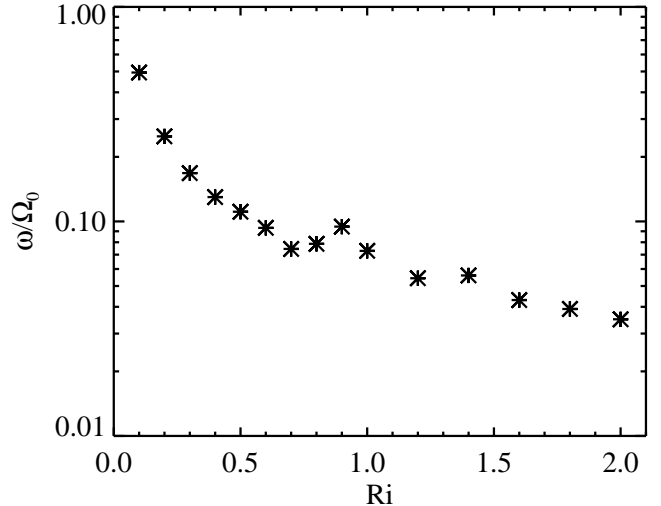}
  \caption{Initial growth rate of the Kelvin-Helmholtz instability as a
      function of the Richardson number ${\rm Ri}$. There is measurable growth
      out to at least ${\rm Ri}=2.0$, which is way beyond the classical value
      of the critical Richardson number of ${\rm Ri}_{\rm c}=0.25$.}
  \label{f:growth}
\end{figure}

The measured growth rates are shown in \Fig{f:growth}. At a Richardson number
close to zero, the growth rate is similar in magnitude to the rotation
frequency $\varOmega_0$ of the disk, whereas for larger values of the
Richardson number, the growth rate falls rapidly. We find that there is growth
out to at least ${\rm Ri}=2.0$, with a growth rate approaching
$\omega=0.01\varOmega_0$. There is no evidence for a cut-off in the growth rate
at the classical value of the critical Richardson number of $0.25$. The range
of unstable Richardson numbers is in good agreement with the mid-plane
Richardson number in the particle simulations shown in
\Figs{f:Rimt_z_t0.02}{f:Rimt_z_t0.1}. This is another confirmation that the
critical Richardson number is around unity or higher when the Coriolis force
is included in the calculations. On the other hand, when
the Keplerian shear is included, growth rates higher than the shear rate $3/2
\varOmega_0$ are expected to be required to overcome the shear
\citep{IshitsuSekiya2003}, although numerical simulations in 3-D would be
required to address these analytical results in detail.

Our measured growth rates are somewhat smaller than in GO05, but we believe
that this is due to the different Coriolis force term in the present work.
Changing the factor $-1/2$ to a factor $-2$ in \Eqs{eq:eqmot2}{eq:eqmotp2}
yields very similar growth rates to GO05. The factor $-1/2$ is a consequence of
the advection of the Keplerian rotation velocity when fluid parcels move
radially, an effect that was not included in the simulations of GO05.

\section{PROPERTIES OF KELVIN-HELMHOLTZ TURBULENCE}

\subsection{Diffusion Coefficient}

The effect of turbulence on the vertical distribution of dust grains can be
quantified as a diffusion process with the turbulent diffusion coefficient
$D_{\rm t}$ \citep{Cuzzi+etal1993,Dubrulle+etal1995,SchraeplerHenning2004}.
Assuming that $D_{\rm t}$ is a constant, i.e.\ independent of the height over
the mid-plane, the equilibrium between vertical settling of dust with velocity
$w_z=-\tau_{\rm f} \varOmega_0^2 z$ and turbulent diffusion implies a vertical
distribution of the dust-to-gas ratio $\epsilon(z)$ that is Gaussian around the
mid-plane \citep{Dubrulle+etal1995},
\begin{equation}
  \epsilon=\epsilon_1 \exp[-z^2/(2 H_\epsilon^2)] \, ,
  \label{eq:eps_strat}
\end{equation}
with the dust-to-gas ratio scale height given by the expression
$H_\epsilon^2=D_{\rm t}/(\tau_{\rm f} \varOmega_0^2)$. Writing now $D_{\rm
t}=\delta_{\rm t} H^2 \varOmega_0$, we get
\begin{equation}\label{eq:deltat}
  \delta_{\rm t} = \left( \frac{H_\epsilon}{H} \right)^2 \varOmega_0\tau_{\rm f}
  \, .
\end{equation}
Using \Eq{eq:deltat}, one can translate the scale-height of the dust-to-gas
ratio $H_\epsilon$ into a turbulent diffusion coefficient $\delta_{\rm t}$.
Such an approach has been used to calculate the turbulent diffusion coefficient
of magnetorotational turbulence \citep{JohansenKlahr2005}. An obvious
difference between Kelvin-Helmholtz turbulence and magnetorotational turbulence
is that dust itself plays the active role for the first, whereas for the latter
the presence of dust does not change the turbulence in any way, because the
local dust-to-gas ratio is assumed to be low. Thus, for Kelvin-Helmholtz
turbulence we expect the diffusion coefficient to depend on the friction time
$\delta_{\rm t}=\delta_{\rm t}(\tau_{\rm f})$.

The calculated turbulent diffusion coefficients for all the simulations are
shown in \Tab{t:results}.  For the dust-to-gas ratio scale height we have, for
simplicity, used the root-mean-square of the $z$-coordinates of all the
particles. The measured coefficients are extremely low, on the order of
$10^{-6}$. If we assume that the turbulent viscosity $\alpha_{\rm t}$ is
similar to the turbulent diffusion coefficient $\delta_{\rm t}$, then one sees
that Kelvin-Helmholtz turbulence is much weaker than magnetorotational
turbulence where $\alpha$-values from $10^{-3}$ to unity are found. There is a
good agreement between the diffusion coefficients of run B and run Br512 (which
has twice the grid and particle resolution). This shows that the solution has
converged and that $256\times128$ is indeed a sufficient resolution to say
something meaningful about the Kelvin-Helmholtz turbulence. For the simulations
with an increased global dust-to-gas ratio (runs Be2, Be5 and Be10), the dust
scale height falls with increasing dust-to-gas ratio. This is to be expected
from \Eq{eq:epsz_Riconst}, because of the cusp of high dust density that forms
around the mid-plane when $\epsilon_1\gg1$. The diffusion coefficient for
$\varOmega_0 \tau_{\rm f}=1.0$ is around $50\%$ lower than for $\varOmega_0
\tau_{\rm f}=0.1$. Here the strong vertical settling of the dust has decreased
the width of the dust layer significantly, and thus the diffusion coefficient
is also much lower. 

\begin{figure}
  \includegraphics{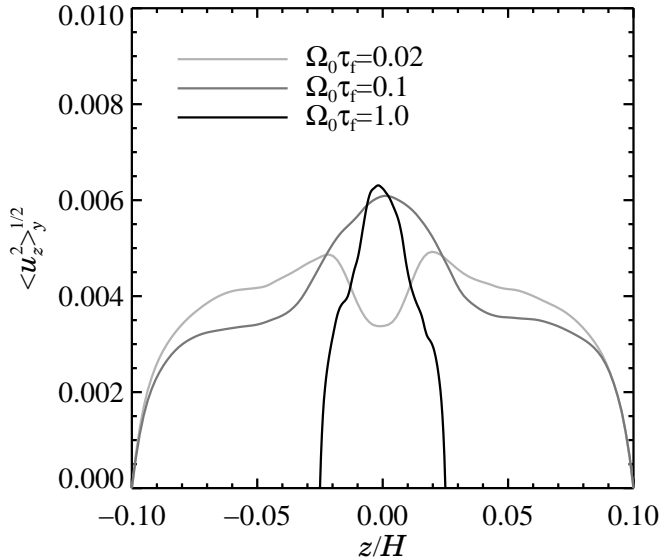}
  \caption{Root-mean-square of the vertical gas velocity as a function of
      height over the mid-plane. The value is quite independent of Stokes
      number, but the width of the turbulent region is very small for
      $\varOmega_0 \tau_{\rm f}=1$ because of the strong vertical settling of
      m-sized boulders. The boundary conditions set the vertical speed to zero
      at the boundaries.}
  \label{f:uzrmsmt_z}
\end{figure}

Turbulent transport coefficients such as $\alpha_{\rm t}$ and $\delta_{\rm t}$
have an inherent dependence on the width of the turbulent region. Thus the
``strength'' of the turbulence is better illustrated by the actual turbulent velocity
fluctuations.
In \Fig{f:uzrmsmt_z} we plot the root-mean-square of the vertical gas velocity
as a function of height over the mid-plane. In the mid-plane, the value is
quite independent of the friction time, whereas the width of the turbulent
region is much smaller for $\varOmega_0 \tau_{\rm f}=1$. Thus the turbulence in
itself is not weaker, only the turbulent region is smaller, and that means that
the transport coefficients are accordingly small.

\subsection{Comparison With Analytical Work}

It is evident from \Tab{t:results} that the diffusion coefficient depends on
the friction time. In the limit of small Stokes numbers, the constant
Richardson number density distribution formulated by S98 predicts that the
vertical dust density distribution should not depend on friction time, and
thus, according to \Eq{eq:deltat}, that the diffusion coefficient should be
proportional to the friction time. The ratio of the diffusion coefficient of
run B to that of run A is $8.9/3.0\approx3$, and not $0.1/0.02=5$ as would give
rise to the same scale height. The factor two difference can be (trivially)
attributed to the slight difference in scale heights for the two runs. The
squared value is different by almost a factor of two, an indication that
vertical settling is not completely negligible for $\varOmega_0 \tau_{\rm
f}=0.1$.
\begin{deluxetable}{lrrr}

  \tablewidth{0pt}
  \tablecaption{Diffusion Coefficients}
  \tablehead{\colhead{Run} & \colhead{$\varOmega_0 \tau_{\rm f}$} &
      \colhead{$\sqrt{\langle z^2\rangle}/H$} &
      \colhead{$\delta_{\rm t}/10^{-6}$}}

  \startdata

    A     & $0.02$ & $0.0121\pm0.0010$ & $3.0\pm0.5$ \\
    B     & $0.10$ & $0.0094\pm0.0008$ & $8.9\pm1.5$ \\
    C     & $1.00$ & $0.0025\pm0.0004$ & $6.5\pm2.4$ \\
    Be2   & $0.10$ & $0.0081\pm0.0006$ & $6.5\pm1.0$ \\
    Be5   & $0.10$ & $0.0047\pm0.0003$ & $2.2\pm0.3$ \\
    Be10  & $0.10$ & $0.0031\pm0.0002$ & $1.0\pm0.1$ \\
    Br512 & $0.10$ & $0.0086\pm0.0005$ & $7.5\pm0.8$

  \enddata

  \tablecomments{Col. (1): Name of run. Col. (2): Stokes number.
      Col. (3): Dust scale height. Col. (4): Diffusion coefficient derived
      from \Eq{eq:deltat}.}
  \label{t:results}
\end{deluxetable}

The strength of the Kelvin-Helmholtz turbulence has also been estimated
analytically by \cite{Cuzzi+etal1993}. They find that the turbulent viscosity
$\nu_{\rm t}$ should be approximately (their equation [21])
\begin{equation}\label{eq:cuzzi_nut}
  \nu_{\rm t} \approx \frac{(\eta v_{\rm K})^2}{\varOmega_0 {\rm Re}^{*2}} \, ,
\end{equation}
where $v_{\rm K}$ is the Keplerian velocity, $\eta$ is the pressure gradient
parameter defined in \Eq{eq:eta_def} and ${\rm Re^{*}}$ is the critical
Reynolds number at which the flow becomes unstable. This value can be
approximated by the Rossby number ${\rm Ro}$, the ratio between the advection
and Coriolis force terms of the flow.  \cite{Dobrovolskis+etal1999} estimate a
value of ${\rm Ro} \approx 20\ldots30$. Using the approximation $\eta \approx
c_{\rm s}^2/v_{\rm K}^2$, \Eq{eq:cuzzi_nut} can be written as
\begin{equation}
  \nu_{\rm t} = D_{\rm t} = \frac{\eta}{{\rm Ro}^2} c_{\rm s}^2 \varOmega_0^{-1} \, ,
\end{equation}
which appears in its dimensionless form simply as
\begin{equation}\label{eq:cuzzi_deltat}
  \delta_{\rm t} = \frac{\eta}{{\rm Ro}^2} \, .
\end{equation}
With $\beta=-0.1$, \Eq{eq:eta_beta} gives $\eta=0.003$. Using ${\rm
Ro}=20\ldots30$, \Eq{eq:cuzzi_deltat} gives $\delta_{\rm t} \sim 3\ldots8
\times 10^{-6}$, quite comparable to the values in \Tab{t:results}. On the
other hand, \Eq{eq:cuzzi_nut} does not produce a dust density distribution with
a constant Richardson number, and would thus greatly overestimate the diffusion
coefficient for even smaller grains. This is also noted by
\cite{CuzziWeidenschilling2005} who constrain the validity of \Eq{eq:cuzzi_nut}
to $\varOmega_0 \tau_{\rm f}>0.01$. Thus our measured diffusion coefficients
are actually in good agreement with both \cite{Cuzzi+etal1993} and with S98.

\section{CONCLUSIONS}\label{ch:conclusions}

The onset of the Kelvin-Helmholtz instability in protoplanetary disks has been
known for decades to be the main obstacle for the formation of planetesimals
via a gravitational collapse of the particle subdisk. Thus the study of the
Kelvin-Helmholtz instability is one of the most intriguing problems of
planetesimal formation. It is also a challenging problem to solve, both
analytically and numerically, because of the coevolution of the two components
gas and dust. Whereas turbulence normally arises from the gas flow alone, in
Kelvin-Helmholtz turbulence the dust grains take the active part as the source
of turbulence by piling up around the mid-plane and thus turning the
energetically favored vertical rotation profile into an unstable shear.
Planetesimal formation would be deceptively simple could the solids only
sediment unhindered, but nature's dislike of thin shear flows precludes this by
making the mid-plane turbulent.

In the current work we have shown numerically that when the dust particles are
free to move relative to the gas, the Kelvin-Helmholtz turbulence acquires an
equilibrium state where the vertical settling of the solids is balanced by the
turbulent diffusion away from the mid-plane. For cm-sized pebbles and dm-sized
rocks, we find that the dust component forms a layer that has a constant
Richardson number. We thus confirm the analytical predictions by
\cite{Sekiya1998} for the first time in numerical simulations.

In the saturated turbulence we find the formation of highly overdense regions
of solids, not in the mid-plane, but embedded in the turbulent flow. The
clumping is very related to the streaming instability found by
\cite{YoudinGoodman2005}. Dust clumps with a density that is equal to or higher
than the gas density orbit at the Keplerian velocity, so the clumps overtake
sub-Keplerian regions of lower dust density.  Thus the dense clumps continue to
grow in size and in mass. The final size of a dust clump is given by a balance
between this feeding and the loss of material in a rarefaction tail that is
formed behind the clump along the sub-Keplerian stream. The gravitational
fragmentation of the single clumps into planetesimals is more likely than the
whole dust layer fragmenting, because the local dust density in the clumps can
be more than an order of magnitude higher than the azimuthally averaged
mid-plane density. This process is very much related to the gravoturbulent
formation of planetesimals in turbulent magnetohydrodynamical flows
\citep*{Johansen+etal2006}.

A full understanding of the role of Kelvin-Helmholtz turbulence in
protoplanetary disks must eventually rely on simulations that include the
effect of the Keplerian shear, so future simulations have to be extended into
three dimensions. One can to first order expect that growth rates of the KHI
larger than the shear rate $\varOmega_0$ are required for a mode to grow in
amplitude faster than it is being sheared out \citep {IshitsuSekiya2003}, but
so far it is an open question in how far the radial shear changes the
appearance of the self-sustained state of Kelvin-Helmholz turbulence. Including
furthermore the self-gravity between the dust particles, it will become
feasible to study the formation of planetesimals in one self-consistent
computer simulation and possibly to answer one of the outstanding questions in
the planet formation process.

\acknowledgments

Computer simulations were performed at the Danish Center for Scientific
Computing in Odense and at the RIO and PIA clusters at the Rechenzentrum
Garching. We would like to thank Andrew Youdin for his critical reading of the
original manuscript. We are also grateful to Gilberto G\'omez for helping to
find the reason for the difference in the KHI growth rates between our
simulations and his own. Our work is supported in part by the European
Community's Human Potential Programme under contract HPRN-CT-2002-00308,
PLANETS.

\end{document}